\documentclass[%
 aip,
 jmp,%
 amsmath,amssymb,
 reprint,%
]{revtex4-1}

\usepackage{color}

\usepackage{graphicx}
\usepackage{dcolumn}
\usepackage{bm}

\begin{document}

\preprint{AIP/123-QED}

\title{Inductance due to spin current}

\author{Wei Chen}
\affiliation{ 
Max Planck Institute for Solid State Research, Heisenbergstrasse 1, D-70569 Stuttgart, Germany
}%

\date{\today}

\begin{abstract}
The inductance of spintronic devices that transport charge neutral spin currents is discussed. It is known that in a media that contains charge neutral spins, a time-varying electric field induces a spin current. We show that since the spin current itself produces an electric field, this implies existence of inductance and electromotive force when the spin current changes with time. The relations between the electromotive force and the corresponding flux, which is a vector calculated by the cross product of electric field and the trajectory of the device, are clarified. The relativistic origin generally renders an extremely small inductance, which indicates the advantage of spin current in building low inductance devices. The same argument also explains the inductance due to electric dipole current, and applies to physical dipoles consist of polarized bound charges.
\end{abstract}

\maketitle


\section{Introduction}

The past few decades witness remarkable progress in the field of spintronics\cite{Wolf01,Zutic04,Fabian07}. Based on transport of spin angular momentum, several principles for generation\cite{Berger96,Slonczewski96,Tserkovnyak02} and detection\cite{Dyakonov71,Hirsch99,Saitoh06,Valenzuela06,Kimura07} of spin current have been clarified or proposed, making building large scale, complex devices approaching reality. It has also been demonstrated that a spin current does not necessarily have to carry charge. Such a charge neutral spin current, or "pure spin current", can be realized in a variety of metallic or insulating systems. For instance, spin waves in magnetic insulators can also transport angular momentum and be converted into the counterpropagate of conduction electrons with opposite spins\cite{Takahashi09,Kajiwara10}. The helical edge states in qusntum spin Hall states\cite{Kane05,Bernevig06,Bernevig06_2,Konig07} and topological superconductors\cite{Schnyder08,Kitaev09,Qi09} also support pure spin currents that consist of counterpropagate of opposite spins. Another example is the spin supercurrent in magnetic insulators that originates from virtual hopping of electrons, typically proportional to the vector chirality ${\bf S}_{i}\times{\bf S}_{j}$ between spins on neighboring sites $i$ and $j$\cite{Hikihara08,Okunishi08,Kolezhuk09,Chen14}, and is shown to be the main mechanism of incommensurate magnetic ordering in multiferroic materials\cite{Katsura05}. Added to the list is the spin Josephson effect between two coupled ferromagnetic metals with misaligned magnetizations\cite{Nogueira04}, whose Josephson current is a pure spin current carried by the magnetic condensate\cite{Chen13,Chen14_2}.

Despite the exciting progress in spintronics, one important aspect regarding building practical devices seems to be omitted in the literature, namely the inductance of spintronic devices. One should be reminded that a motive that drives the research in spintronic devices is their advantage over electronic devices, such as lower power consumption. However, as far as inductance is concerned, it is unclear at present if spintronic devices, especially those transport pure spin currents, are still better options compared to electronic devices.

In this article, we clarify the origin of the inductance due to pure spin currents, discussed entirely within the framework of classical electromagnetism. We noticed that, following the discovery of Aharonov-Casher (AC) effect\cite{Aharonov84}, it has been recognized that a time-varying electric (${\bf E}$) field renders a spin motive force on spin $1/2$ particles, which can be formulated within a $SU(2)$ gauge theory that treats the spin as gauge charge\cite{Anandan89,Casella92,Lee94,Ryu96}. Owing to this spin motive force, we show that when a spin current changes with time, the Lorentz boost of the spins creates a time-varying ${\bf E}$ field that in turn exerts a force on the spins, hence the self-inductance. The induction is related to a proposed "electric flux vector" calculated by the cross product of the electric field and the particle trajectory. The relation between spin motive force, inductance, and electric flux vector bears a form similar to Faraday's law of induction\cite{Ryu96}. The analog of Lenz's law can be deduced from the energy conservation argument \cite{Griffith89}, which states that the induced emf always opposes changing of the spin current. Moreover, the electric flux vector is consistent with that defined from quantum interference of magnetic dipoles \cite{Chen13}. This correspondence clarifies the fundamental role of the electric flux vector in both quantum mechanics and classical electromagnetism. We further show that because of the relativistic origin, the inductance due to spin current is practically negligible, hence pure spin currents are indeed better options over electric currents for devices that require low inductance. On the other hand, in low dimensional systems where the relativistic effect is generally enhanced by several orders of magnitude, the emf due to a time-varying electric field may create small yet detectable pure spin currents.

Besides the inductance of spintronic devices, the duality between electricity and magnetism immediately suggests that the inductance of devices that transport electric dipole currents also has the same relativistic origin. The analog of Faraday's law in this case is related to a "magnetic flux vector" defined similarly to the electric flux vector. Although there is no sizable point electric dipoles in nature, we show that the same principle also applies to physical dipoles consists of electrons and holes bound by Coulomb interaction, as recently realized in semiconductor heterostructures\cite{Spielman00,Kellogg04,Tutuc04,Eisenstein04,Tiemann08,Huang12,Sivan92,Seamons07,DasGupta08} and bilayer graphene\cite{Kim11,Kim12,Gorbachev12,Berman12}, as well as dipolar molecules. A time variance of magnetic flux vector can generate small yet detectable electric dipole currents as long as the system has strong relativistic effect.

\section{Inductance due to spin current}

To discuss the self-inductance due to dipole currents, we begin by considering the force on a magnetic dipole moving in an ${\bf E}$ field, calculated from the relativistic Lagrangian \cite{Aharonov84} (SI units are adopted through out the article. Bold face symbols are vectors, except inductance tensor ${\bf L}_{\mu,d}$ defined later)
\begin{eqnarray}
{\cal L}_{\mu}=\frac{1}{2}m{\bf v}\cdot{\bf v}+\frac{1}{c^{2}}{\bf E}\cdot\left({\bf v}\times{\boldsymbol \mu}\right)
\label{spin_Lagrangian}
\end{eqnarray}
The second term comes from the Lorentz boost of the magnetic dipole, which yields a electric dipole \cite{Hirsch90} ${\bf d}={\bf v}\times{\boldsymbol\mu}/c^{2}$ that interacts with external electric field in the lab frame (we only take the leading order in ${\bf v}/c$ by assuming $|{\bf v}|\ll c$). The force calculated from classical equation of motion is
\begin{eqnarray}
{\bf F}=\frac{d}{dt}m{\bf v}&=&
\frac{1}{c^{2}}{\boldsymbol\nabla}\left[{\bf E}\cdot\left({\bf v}\times{\boldsymbol\mu}\right)\right]
-\frac{1}{c^{2}}{\boldsymbol\mu}\times\frac{ d{\bf E}}{ d t}
\label{force_general}
\end{eqnarray}
which recovers the same spin motive force (to leading order in ${\bf E}$) derived rigorously from covariant derivative of the $SU(2)$ gauge theory for spinful particles\cite{Anandan89,Casella92,Lee94,Ryu96}. Another support of this classical formalism is that it derives the Rashba spin-orbit coupling (SOC) in condensed matter systems where Lorentz invariance is generally irrelevant. Either from this classical formalism or the $SU(2)$ gauge theory, we stress that our point is not to repeat the detail derivation of the spin motive force, but to show that it is intimately related to the inductance due to spin currents. Moreover, Eq. (2) implies that the spin motive force also manifests in spintronic devices which operate at much lower energy scale that Lorentz invariance is unimportant. Notice that we express the force in terms of $\frac{d}{dt}=\frac{\partial}{\partial t}+{\bf v}\cdot{\boldsymbol\nabla}$ instead of $\frac{\partial}{\partial t}$ as in most of the literature\cite{Anandan89,Lee94}, and ignored the effect of magnetic field, both for the sake of discussing inductance. If the dipole is moving in a closed trajectory, the work done by the force is 
\begin{eqnarray}
W=-\frac{1}{c^{2}}\oint{\boldsymbol\mu}\times\frac{ d{\bf E}}{ d t}\cdot d{\bf l}=-\frac{1}{c^{2}}{\boldsymbol\mu}\cdot\left(\oint \frac{ d {\bf E}}{ d t}\times d{\bf l}\right)
\label{work_spin_force}
\end{eqnarray}
by assuming a constant ${\boldsymbol\mu}$. The gradient term in Eq.~(\ref{force_general}) drops out after loop integration.

The dc spin current generated by the spin motive force can be described by a Drude model. It is, however, unlikely to create such a dc current over a long period of time, since a constant $d{\bf E}/dt$ requires increasing ${\bf E}$ with time indefinitely. The ac current generated by a homogeneous, time varying ${\bf E}(t)=Re({\bf E}(\omega)e^{-i\omega t})$ is more promising, which yields an ac conductivity $\sigma_{\mu}(\omega)=n|{\boldsymbol\mu}|^{2}\tau/m(1-i\omega\tau)$, where $\tau$ is the mean free time. Despite its similarity to the usual ac conductivity, an important difference is that Drude model is adequate only if the time scale at which ${\bf E}(t)$ varies is much longer than $\tau$, such that many collision events happen within the time period of the driving force. This constraints the frequency of the driving force $\omega\tau\ll 1$.

We now address self-inductance. Wires carrying a spin current have self-inductance simply because {\it the spin current itself produces electric field} \cite{Hirsch90,Hirsch99,Meier03,Schutz03,Sun04}. Consider the case that changing of electric field $ d{\bf E}/ d t$ is due to changing of the spin current itself $dI_{\mu}/dt$. One can always write
\begin{eqnarray}
\oint \frac{ d {\bf E}}{ d t}\times d{\bf l}=c^{2}{\bf L}_{\mu}\cdot{\boldsymbol{\hat \mu}}\frac{ d I_{\mu}}{ d t}
\label{L_definition}
\end{eqnarray}
which means after integrating $ d {\bf E}/ d t\times d{\bf l}$ along the wire, the result is a quantity that points at the direction ${\bf L}_{\mu}\cdot{\boldsymbol {\hat \mu}}$ and proportional to $dI_{m}/dt$. Here ${\bf L}_{\mu}$ represents self-inductance tensor that depends on the shape of the wire, and ${\boldsymbol {\hat \mu}}$ is the unit vector along the direction of ${\boldsymbol\mu}$. Writing Eq. (\ref{work_spin_force}) in terms of Eq. (\ref{L_definition}), one obtains a spin motive force
\begin{eqnarray}
{\cal E}_{\mu}&=&\frac{W}{|{\boldsymbol \mu}|}=-{\boldsymbol{\hat \mu}}\cdot{\bf L}_{\mu}\cdot {\boldsymbol {\hat\mu}}\frac{ d I_{\mu}}{ d t}\;.
\label{self_inductance}
\end{eqnarray}
This means a changing spin current induces a spin motive force ${\cal E}_{m}$ that tends to oppose the change of spin current, which is precisely the meaning of self-inductance. Eq. (\ref{L_definition}) suggests that the flux caused by the spin current itself is a vector 
\begin{eqnarray}
{\bf \Phi}_{E}=\oint d{\bf l}\times{\bf E}=c^{2}{\bf L}_{\mu}\cdot{\boldsymbol{\hat \mu}}I_{m}\;,
\label{Phi_definition}
\end{eqnarray}
and is related to the self-inductance and spin current by ${\cal E}_{\mu}=-{\boldsymbol {\hat\mu}}\cdot d{\boldsymbol \Phi}_{E}/c^{2}dt$,  analogous to Faraday's law of induction. Compared to previous investigations on spin motive force\cite{Ryu96} that write ${\cal E}=-\left(\hbar/|{\boldsymbol\mu}|\right)d\Phi^{AC}/dt$ with $\Phi^{AC}={\boldsymbol\mu}\cdot{\boldsymbol\Phi}_{E}/\hbar c^{2}$, the expressions in Eqs.~(\ref{self_inductance}) and (\ref{Phi_definition}) clarify the flux caused by the spin current itself, and the role of inductance that is only determined by the shape of the wire. We remark that ${\boldsymbol\Phi}_{E}$ also manifests itself in quantum mechanics, where it characterizes the phase acquired by a magnetic dipole moving in an ${\bf E}$ field \cite{Chen13}.

\begin{figure}[ht]
\begin{center}
\includegraphics[clip=true,width=0.99\columnwidth]{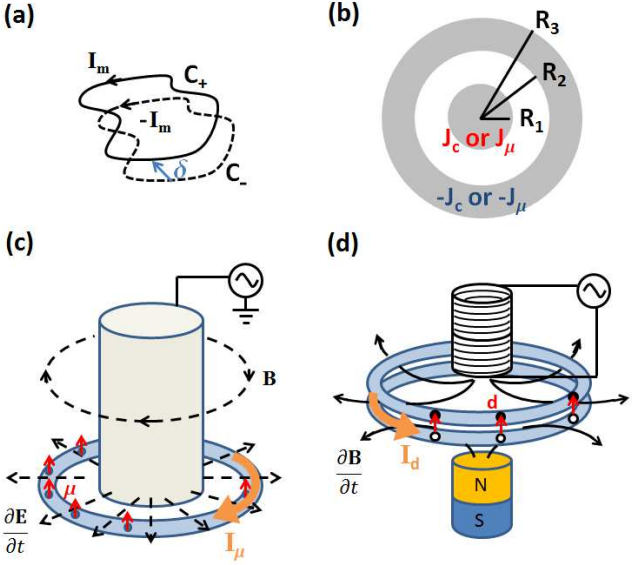}
\caption{ (color online) (a) Schematics of decomposing a spin current into two fictitious monopole currents $\pm I_{m}$ running on two parallel trajectories $C_{\pm}$. (b) The coaxial cable for comparing inductance $L_{c}$ and ${\boldsymbol{\hat \mu}}\cdot{\bf L}_{\mu}\cdot {\boldsymbol {\hat\mu}}$. The two gray areas carry charge current or spin current in the opposite direction, with radii $R_{1}\sim R_{3}$ of the same order. (c)Proposed device to generate pure spin currents by applying $\partial{\bf E}/\partial t$ on spin polarized excitons. (d)Proposed device to generate electric dipole current by applying $\partial{\bf B}/\partial t$ on bilayer excitons. }
\label{fig:infinite_wire}
\end{center}
\end{figure}

To claim that ${\cal E}_{\mu}$ in Eq.~(\ref{self_inductance}) acts {\it against} changing of spin current, i.e., Lenz's law, one needs to prove ${\boldsymbol{\hat \mu}}\cdot{\bf L}_{\mu}\cdot {\boldsymbol {\hat\mu}}>0$ holds for wires of any shape and with any orientation of magnetization ${\boldsymbol {\hat\mu}}$. To see this, we introduced two fictitious monopole currents that compose the spin current \cite{Nagaosa08}. Consider $+I_{m}$ and $-I_{m}$ running on two closed paths, denoted by $C_{+}$ and $C_{-}$, that are of the same shape and lie on parallel planes that are distance ${\boldsymbol\delta}$ apart, as shown in Fig. \ref{fig:infinite_wire}(a). The $+q_{m}$ at ${\bf r}_{+}$ on $C_{+}$ is bound to the $-q_{m}$ at ${\bf r}_{-}={\bf r}_{+}-{\boldsymbol\delta}$ on $C_{-}$. The path spanned by ${\bf r}=\left({\bf r}_{+}+{\bf r}_{-}\right)/2$ is trace of the spin current denoted by $C_{0}$. We denote the electric field produced by $+I_{m}$ running on $C_{0}$ as ${\bf E}_{0}$. Now the field experienced on $C_{0}$ from the $+I_{m}(-I_{m})$ running on $C_{+}(C_{-})$ is $\pm{\bf E}_{0}-\left(\frac{{\boldsymbol\delta}}{2}\cdot{\boldsymbol\nabla}\right){\bf E}_{0}$ for small $|{\boldsymbol\delta}|$, so the total field on $C_{0}$ is 
\begin{eqnarray}
{\bf E}_{dip}=-({\boldsymbol\delta}\cdot{\boldsymbol\nabla}){\bf E}_{0}\;.
\end{eqnarray}
The work done by external power source against ${\cal E}_{\mu}$ to build up $I_{\mu}$ is 
\begin{eqnarray}
W^{ext}&=&\frac{1}{2}{\boldsymbol{\hat\mu}}\cdot{\bf L}_{\mu}\cdot{\boldsymbol{\hat\mu}}I_{\mu}^{2}
=\frac{1}{2c^{2}}I_{\mu}{\boldsymbol{\hat\mu}}\cdot\oint_{C_{0}}{\bf E}_{dip}\times d{\bf l}
\nonumber \\
&=&\frac{1}{2c^{2}}{\boldsymbol\delta}\cdot\int{\bf E}_{dip}\times{\bf J}_{m}d\tau
\label{work_spin_1st_step}
\end{eqnarray}
where we have used $I_{\mu}{\boldsymbol{\hat\mu}}=I_{m}{\boldsymbol\delta}$, and $\oint I_{m}d{\bf l}=\int {\bf J}_{m}d\tau$. Since ${\boldsymbol\nabla}\times{\bf E}_{0}=-\mu_{0}{\bf J}_{m}$, 
\begin{eqnarray}
W^{ext}&=&-\frac{\epsilon_{0}}{2}\int{\bf E}_{dip}\cdot\left[\left({\boldsymbol\nabla}\times{\bf E}_{0}\right)\times{\delta}\right]d\tau
\nonumber \\
&=&\frac{\epsilon_{0}}{2}\int d\tau|{\bf E}_{dip}|^{2}+\frac{\epsilon_{0}}{2}\int d{\bf a}\cdot\left[\left({\boldsymbol\delta}\cdot{\bf E}_{0}\right){\bf E}_{dip}\right]\;,
\label{work_spin_2nd_step}
\end{eqnarray}
where we have used two vector identities ${\boldsymbol\nabla}\left({\boldsymbol\delta}\cdot{\bf E}_{0}\right)={\boldsymbol\delta}\times\left({\boldsymbol\nabla}\times{\bf E}_{0}\right)+\left({\boldsymbol\delta}\cdot{\boldsymbol\nabla}\right){\bf E}_{0}$ for constant ${\boldsymbol\delta}$, and ${\bf E}_{dip}\cdot\left[{\boldsymbol\nabla}\left({\boldsymbol\delta}\cdot{\bf E}_{0}\right)\right]={\boldsymbol\nabla}\cdot\left[\left({\boldsymbol\delta}\cdot{\bf E}_{0}\right){\bf E}_{dip}\right]-\left({\boldsymbol\delta}\cdot{\bf E}_{0}\right){\boldsymbol\nabla}\cdot{\bf E}_{dip}$, and the fact that ${\boldsymbol\nabla}\cdot{\bf E}_{dip}=-\left({\boldsymbol\delta}\cdot{\boldsymbol\nabla}\right)\left({\boldsymbol\nabla}\cdot{\bf E}_{0}\right)=0$ since there is no charge in this problem. The surface term in Eq. (\ref{work_spin_2nd_step}) drops out at $r\rightarrow\infty$. We conclude that 
\begin{eqnarray}
W^{ext}&=&{\boldsymbol{\hat\mu}}\cdot{\bf L}_{\mu}\cdot{\boldsymbol{\hat\mu}}I_{\mu}^{2}/2
\nonumber \\
&=&\epsilon_{0}\int d\tau|{\bf E}_{dip}|^{2}/2=W^{field}>0\;,
\end{eqnarray}
so the work is indeed stored as the energy of the dipolar field \cite{Griffith89}, and ${\boldsymbol{\hat\mu}}\cdot{\bf L}_{\mu}\cdot{\boldsymbol{\hat\mu}}>0$, so Lenz's law is justified.

The magnitude of ${\boldsymbol{\hat \mu}}\cdot{\bf L}_{\mu}\cdot {\boldsymbol {\hat\mu}}$, in comparison with the usual charge inductance $L_{c}$, can be estimated by considering an infinitely long coaxial cable with finite thickness, as shown in Fig.~\ref{fig:infinite_wire}(b). The radii $R_{1}\sim R_{3}$ are of the same order. We compare a charge current $J_{c}=\pm nev$ running on inner and outer part of the coaxial cable with spin current $J_{\mu}=\pm n\mu_{B}v$ that has the same particle density $n$ and velocity $v$. This comparison is meaningful in the sense that the two currents transport the same rate of particles, but one carries electron charge whereas the other carries Bohr magneton. For the charge current, the work per unit length against the inductance stores in the magnetic field $W^{B}/l=L_{c}I_{c}^{2}/2l\sim\mu_{0}J^{2}R_{1}^{4}\times {\cal O}(1)$, and inductance per unit length is $L_{c}/l\sim\mu_{0}$. For the spin current case the work stores in the electric field 
\begin{eqnarray}
W^{E}/l={\boldsymbol{\hat \mu}}\cdot{\bf L}_{\mu}\cdot {\boldsymbol {\hat\mu}}I_{m}^{2}/2l\sim\mu_{0}^{2}\epsilon_{0}J_{m}^{2}R_{1}^{2}\times{\cal O}(1)\;,
\end{eqnarray}
so the spin inductance per unit length ${\boldsymbol{\hat \mu}}\cdot{\bf L}_{\mu}\cdot {\boldsymbol {\hat\mu}}/l\sim\mu_{0}^{2}\epsilon_{0}/R_{1}^{2}$ depends on the width of the coaxial cable. The ratio of the two works is 
\begin{eqnarray}
W^{E}/W^{B}\sim \mu_{0}\epsilon_{0}|{\boldsymbol \mu}|^{2}/e^{2}R_{1}^{2}\;.
\end{eqnarray}
For a wire of mm size one obtains $W^{E}/W^{B}\sim 10^{-19}$. The smallness of ${\boldsymbol{\hat \mu}}\cdot{\bf L}_{\mu}\cdot {\boldsymbol {\hat\mu}}$ is not surprising because of its relativistic origin. Small inductance implies that to detect a spin current by inductance, as in the design that that uses the charge inductance of a pick up coil to detect the charge supercurrent in an rf superconducting quantum interference device (SQUID), is practically not feasible. On the other hand, small inductance also has advantages in practical devices. For instance, in an $RL$ circuit, it takes much shorter time to saturate a spin current than a charge current. Evidently, our analysis also implies that if a spintronic device also transports charge, such as those operated with spin-polarized electrons, then its inductance almost entirely comes from the charge current.

Previous investigations on spin motive force concern particles that preserve Lorentz invariance. It is, however, intriguing to discuss the feasibility of generating spin current by the spin motive force in solid state devices, derived in Eq.~(\ref{force_general}) from classical equation of motion. There are several options for charge neutral magnetic dipoles, for instance liquid $^{3}$He and ferrofluids. But as far as building practical devices is concerned, spin polarized excitons in semiconductors seem to be more promising\cite{Rashba82}. As sketched in Fig. \ref{fig:infinite_wire}(c), we proposed that by applying $\partial{\bf E}/\partial t$ in a semiconducting ring that contains spin polarized excitons generated by polarized light \cite{Rashba82}, the spin current $I_{m}$ can be realized. To optimize the current, ${\boldsymbol\mu}\parallel{\bf \hat{z}}$ and ${\bf E}(t)={\bf E}_{0}\cos\omega t$ with ${\bf E}_{0}\parallel{\bf {\hat r}}$ should be achieved. As discussed previously, $1/\omega$ should be longer than $\tau$ such that Drude model is adequate, and ${\bf E}_{0}$ should be smaller than the field ionization scale which is typically $\sim 10^{6}$V/m for excitons \cite{Miller85}. Assuming $\tau\sim 10^{-15}$s can be achieved, an appropriate ac field is that produced by a metallic cylinder sheet with radius $R$ and oscillating uniform surface charge density $\sigma(t)=\sigma_{0}\cos\omega t$, which yields ${\bf E}={\bf {\hat r}}\left(\sigma_{0}R/\epsilon_{0}r\right)\cos\omega t$, as sketched in Fig. \ref{fig:infinite_wire}(c). For concreteness, we consider $|{\bf E}|\sim 10^{3}$V/m and $\omega\sim 10^{13}$Hz in this setup. The changing ${\bf E}$ field yields a force $|{\bf F}|\sim 10^{-24}$N according to Eq. (\ref{force_general}), and a velocity $|{\bf v}|\sim 10^{-8}$m/s within Drude model for an exciton with mass $m\sim 0.1m_{e}$. Although this velocity seems negligible, it can be dramatically enhanced in systems that have strong Rashba SOC. For instance, in semiconductor heterostructures the Rashba SOC is equivalent to replacing $1/c^{2}$ in Eqs. (\ref{spin_Lagrangian})$\sim$(\ref{Phi_definition}) by a phenomenological prefactor $\lambda$ that can be $5$ orders of magnitude larger\cite{Chen13}. The drifting velocity $|{\bf v}|\sim 10^{-3}$m/s increases accordingly in the ac field considered here. Thus we anticipate that a strong Rashba SOC is necessary to produce a observable spin current via time-varying ${\bf E}$ field.

The changing ${\bf E}$ field also induces a magnetic field ${\bf B}={\boldsymbol{\hat\phi}}\left(\mu_{0}\sigma_{0}\omega zR/r\right)\sin\omega t$. This is simply the magnetic field due to the oscillating current that pumps the surface charge in and out of the cylinder. Since it is linear to $z$, one can put the semiconducting ring at the bottom of the cylinder(we ignore any complication at the boundary of the cylinder) to minimize ${\bf B}$, so its effect on ${\boldsymbol\mu}$, such as Larmor precession or torque, can be ignored, and the assumption of fixed ${\boldsymbol \mu}$ is satisfied. The exciton current may be measured by verifying the Joule heating law.

The force in Eq.~(\ref{force_general}) can serve as the driving force for excitons to exhibit other phenomena. An example of particular interest at present is the spin Hall effect (SHE) of excitons\cite{Kuga08}. If the excitons experience Berry curvature in momentum space, the anomalous velocity causes spin up and down excitons to deflect into opposite directions, if they are accelerate by certain force at the first place. For the set up in Fig.~\ref{fig:infinite_wire} (c) that contains strong Rashba SOC, the force can reach $F\sim 10^{-19}$N or higher, comparable to that produced by an uniaxial strain originally proposed to accelerate the excitons\cite{Kuga08}. Thus a time-varying ${\bf E}$ field may be an efficient way to accelerate excitons and observe SHE, although the ac nature of the force also constraints the time scale and distance that excitons travel.

\begin{table}
\begin{tabular}{p{1.5cm}<{\centering}  p{3.5cm}<{\centering}  p{3.5cm}<{\centering} }
\hline \hline

     & electric & magnetic \\ \hline

monopole & $\Phi_{B}=\int d{\bf a}\cdot{\bf B}$  & $\Phi_{E}=-\int d{\bf a}\cdot{\bf E}$ \\ 

 & $=L_{c}I_{c},$ & $=c^{2}L_{m}I_{m},$ \\

 & ${\cal E}_{c}=- d\Phi_{B}/ d t$ & ${\cal E}_{m}=- d\Phi_{E}/c^{2} d t$ \\

 & $=-L_{c} d I_{c}/ d t.$ & $=-L_{m} d I_{m}/ d t.$ \\ \hline

dipole & ${\boldsymbol\Phi}_{B}=\oint d{\bf l}\times{\bf B}$ & ${\boldsymbol\Phi}_{E}=-\oint d{\bf l}\times{\bf E}$  \\ 

 & $={\bf L}_{d}\cdot{\bf{\hat d}}I_{d},$ &  $=c^{2}{\bf L}_{\mu}\cdot{\boldsymbol{\hat \mu}}I_{\mu},$ \\

 & ${\cal E}_{d}= -{\bf {\hat d}}\cdot d{\boldsymbol \Phi}_{B}/ d t$ &  ${\cal E}_{\mu}= -{\boldsymbol {\hat\mu}}\cdot d{\boldsymbol \Phi}_{E}/c^{2} d t$ \\

 & $=-{\bf {\hat d}}\cdot {\bf L}_{d}\cdot{\bf {\hat d}} d I_{d}/ d t.$ &  $=-{\boldsymbol {\hat\mu}}\cdot {\bf L}_{\mu}\cdot{\boldsymbol{\hat \mu}} d I_{\mu}/ d t.$ \\ 

\hline \hline

\end{tabular}	
\caption{ Relations between flux, inductance, and motive force, classified according to the electric or magnetic monopole or dipole moment of the particles. }
\label{tbl:summary_inductance}
\end{table}

\section{Inductance due to electric dipole current}

The analysis for spin current also applies to electric dipole current. Starting from the relativistic Lagrangian for an electric dipole moving in a magnetic field \cite{Wilkens94} 
\begin{eqnarray}
{\cal L}_{d}=\frac{1}{2}m{\bf v}\cdot{\bf v}+{\bf B}\cdot\left({\bf d}\times{\bf v}\right)
\label{Lagrangian_electric_dipole}
\end{eqnarray}
The last term again comes from Lorentz boost of the electric dipole. Following Eq.~(\ref{force_general}), the motive force can be derived from classical equation of motion without concerning Lorentz invariance. For dipoles moving on a closed trajectory, the motive force integrated along the loop is
\begin{eqnarray}
W_{d}=-\oint\frac{ d{\bf B}}{ d t}\times{\bf d}\cdot d{\bf l}=-{\bf d}\cdot\left(\oint d{\bf l}\times\frac{ d {\bf B}}{ d t}\right)
\label{work_electric_dipole_force}
\end{eqnarray}
Since an electric dipole current itself produces magnetic field, the existence of inductance is evident. The inductance tensor is defined by 
\begin{eqnarray}
\oint d{\bf l}\times \frac{d{\bf B}}{dt}={\bf L}_{d}\cdot{\bf {\hat d}}\frac{dI_{d}}{dt}\;.
\end{eqnarray}
The corresponding magnetic flux vector is 
\begin{eqnarray}
{\boldsymbol \Phi}_{B}=\oint d{\bf l}\times{\bf B}={\bf L}_{d}\cdot{\bf {\hat d}}I_{d}\;,
\end{eqnarray}
and hence ${\cal E}_{d}=-{\bf{\hat d}}\cdot d{\boldsymbol\Phi}_{B}/dt$. Introducing two charge currents that compose the electric dipole current and following the procedure in Eqs. (\ref{work_spin_1st_step})$\sim$(\ref{work_spin_2nd_step}), one can also verify Lenz's law ${\bf {\hat d}}\cdot{\bf L}_{d}\cdot{\bf{\hat d}}>0$.

There is no sizable point ${\bf d}$ in nature, but Eq. (\ref{work_electric_dipole_force}) also applies to a physical dipole composed of $\pm q$ that are ${\boldsymbol\delta}$ apart, as we prove below. Consider the geometry shown in Fig. \ref{fig:infinite_wire}(a) again, with particles carrying $+q$($-q$) situated in $C_{+}$($C_{-}$). Each $-q$ is rigidly bound to a $+q$ that is ${\boldsymbol\delta}$ away. In the presence of a time varying ${\bf B}$ field, the force on $C_{\pm}$ is ${\bf F}_{\pm}=\mp qd{\bf A}({\bf r}_{\pm})/dt$, so the work done on a pair of $\pm q$ is, for small $|{\boldsymbol\delta}|$,
\begin{eqnarray}
&&\sum_{\sigma=\pm}\oint_{C_{\sigma}}{\bf F}_{\sigma}\cdot d{\bf r}_{\sigma}=-\frac{d}{dt}q\oint_{C_{0}}\left({\boldsymbol\delta}\cdot{\boldsymbol\nabla}\right){\bf A}\cdot d{\bf r}
\nonumber \\
&&=\frac{d}{dt}q\oint_{C_{0}}{\boldsymbol\delta}\times\left({\boldsymbol\nabla}\times{\bf A}\right)\cdot d{\bf r}=W_{d}\;,\;
\label{work_two_charge}
\end{eqnarray}
where we have used $\left({\boldsymbol\delta}\cdot{\boldsymbol\nabla}\right){\bf A}={\boldsymbol\nabla}\left({\boldsymbol\delta}\cdot{\bf A}\right)-{\boldsymbol\delta}\times\left({\boldsymbol\nabla}\times{\bf A}\right)$ and $\oint{\boldsymbol\nabla}\left({\boldsymbol\delta}\cdot{\bf A}\right)\cdot d{\bf r}=0$. From Eqs. (\ref{work_electric_dipole_force}) and (\ref{work_two_charge}), is it clear that $d{\bf B}/dt$ also creates dipole current in a media composed of physical dipoles. The inductance per unit length of the coaxial cable in Fig. \ref{fig:infinite_wire}(b) carrying electric dipole current $J_{d}=\pm n|{\bf d}|v$ can be estimated from the energy stored in the magnetic field it produces, which yields ${\bf {\hat d}}\cdot{\bf L}_{d}\cdot{\bf {\hat d}}/l\sim \mu_{0}/R_{1}^{2}$. Comparing with a charge current that transports the same amount of particles $J_{c}=\pm nev$, the ratio of time constants in an $RL$ circuit is $\tau_{RL}^{d}/\tau_{RL}^{c}=|{\bf d}|^{2}/e^{2}R_{1}^{2}=|{\boldsymbol\delta}|^{2}/R_{1}^{2}\ll 1$, so $J_{d}$ is also a much better choice than $J_{c}$ as far as inductance is concerned. 

Besides electrically polarized atoms or molecules, a likely candidate for sizable physical dipole is the bilayer exciton that binds an electron and a hole that reside in different planes \cite{Spielman00,Kellogg04,Tutuc04,Eisenstein04,Tiemann08,Huang12,Sivan92,Seamons07,DasGupta08,Kim11,Kim12,Gorbachev12,Berman12}. At present, $|{\boldsymbol\delta}|\sim 10^{-8}$m can be achieved. We assume the dipoles remain rigid at temperature $k_{B}T\ll qV$ smaller than the binding energy. In the device proposed in Fig. \ref{fig:infinite_wire}(d), the ac current in a solenoid produces $\partial{\bf B}/\partial t=\omega{\bf B}_{0}\cos\omega t$. For $\omega\sim 10^{9}$Hz, $|{\bf B}_{0}|\sim 1$G, $\tau\sim 10^{-15}$s, and $m\sim 0.1m_{e}$, this yields a drifting velocity $\sim 10^{-6}$m/s. The magnetic field may cause precession on the spin of the excitons, but this does not contribute to the electric dipole current.

Finally, merely for the sake of completing the duality between electricity and magnetism, and between monopole moments and dipole moments, we address the inductance due to magnetic monopole currents, deduced from Maxwell's equations and Lorentz force that contain monopole terms. Consider ideal, classical magnetic monopoles $q_{m}$ in a closed ring, and a electric field pierced through the ring. Time variance of electric flux $\Phi_{E}=-\int {\bf E}\cdot d{\bf a}$ induces a motive force\cite{Jackson99} $-d\Phi_{E}/c^{2}dt=\oint{\bf F}\cdot d{\bf l}/q_{m}={\cal E}_{m}$, where ${\bf F}=q_{m}\left({\bf B}-{\bf v}\times{\bf E}/c^{2}\right)$ is the Lorentz force experienced by a monopole. A monopole current itself produces electric field according to ${\boldsymbol\nabla}\times{\bf E}=-\mu_{0}{\bf J}_{m}$, where ${\bf J}_{m}$ is the monopole current density. Hence a stable monopole current $I_{m}$ creates a flux that can always be written as $\Phi_{E}=c^{2}L_{m}I_{m}$, where $L_{m}$ is the inductance. Changing $I_{m}$ then induces a motive force ${\cal E}_{m}=-L_{m} d I_{m}/ d t$. Following the energy conservation argument\cite{Griffith89}, Lenz's law can be justified. We stress that this analysis does not apply to the emerged monopoles in spin ice\cite{Castelnovo08,Morris09,Bramwell09,Ryzhkin05}, since the monopoles therein do not modify Maxwell's equations and experience Lorentz force. Nevertheless, this analysis completes the duality summarized in Table \ref{tbl:summary_inductance}.


\section{Conclusions}

In summary, starting from the well known spin motive force in the presence of a time-varying ${\bf E}$ field, we show that since a pure spin current itself produces a electric field, the existence of inductance due to spin current is evident. An energy conservation argument is provided to prove Lenz's law, i.e., the spin motive force due to inductance always opposes changing the spin current. Together with its manifestation in quantum interference of magnetic dipoles\cite{Chen13}, the fundamental roles of the electric flux vector, defined as the cross product of ${\bf E}$ field and trajectory, in both quantum mechanics and classical electromagnetism are clarified.

The significance of our calculation is that, on one hand we prove that the inductance of spintronic devices is practically negligible owing to its relativistic origin, hence they are better options for devices requiring small inductance. On the other hand, we show that in systems with strong Rashba SOC, such as semiconductor heterostructures, the inductance is several orders of magnitude enhanced, hence it is possible to generate a detectable spin current via time-varying electric field. Finally, from the duality between electricity and magnetism, it is obvious that the inductance due to electric dipole currents also has the same relativistic origin. A time-varying magnetic field can generate a detectable electric dipole current in systems that display strong relativistic effect, such as low-dimensional systems that contain polarized bound charges.

\begin{acknowledgments}
We thank G. S. Uhrig, O. P. Sushkov, P. Horsch, J. H. Smet, and R. Zeyher for stimulating discussions.
\end{acknowledgments}


\begin{thebibliography}{99}

\bibitem{Wolf01}
S. A. Wolf {\it et al.}, Science {\bf 294}, 1488 (2001).

\bibitem{Zutic04}
I. \v{Z}uti\'{c}, J. Fabian, and S. Das Sarma, Rev. Mod. Phys. {\bf 76}, 323 (2004).

\bibitem{Fabian07}
J. Fabian, A. Matos-Abiague, C. Ertler, P. Stano, and I. \v{Z}uti\'{c}, Acta Phys. Slov. {\bf 57}, 565 (2007).

\bibitem{Berger96}
L. Berger, Phys. Rev. B {\bf 54}, 9353 (1996).

\bibitem{Slonczewski96}
J. C. Slonczewski, J. Magn. Magn. Mater. 159, L1-L7
(1996).

\bibitem{Tserkovnyak02}
Y. Tserkovnyak, A. Brataas, and G. E. W. Bauer, Phys. Rev. Lett. {\bf 88}, 117601 (2002).

\bibitem{Dyakonov71}
 M. I. Dyakonov and V. I. Perel, JETP Lett. {\bf 13}, 467 (1971); Phys. Lett. {\bf 35A}, 459 (1971).

\bibitem{Hirsch99}
J. E. Hirsch, Phys. Rev. Lett. 83, 1834 (1999).

\bibitem{Saitoh06}
E. Saitoh {\it et al.}, Appl. Phys. Lett. {\bf 88}, 182509 (2006).
 
\bibitem{Valenzuela06}
S. O. Valenzuela and M. Tinkham, Nature {\bf 442}, 176 (2006). 

\bibitem{Kimura07}
T. Kimura {\it et al.}, Phys. Rev. Lett. {\bf 98}, 156601 (2007). 

\bibitem{Takahashi09}
S. Takahashi, E. Saitoh, and S. Maekawa, J. Phys. Conf. Ser. {\bf 200}, 062030 (2009).

\bibitem{Kajiwara10}
Y. Kajiwara {\it et al.}, Nature {\bf 464}, 262 (2010).


\bibitem{Kane05}
C. L. Kane and E. J. Mele, Phys. Rev. Lett. {\bf 95}, 226801 (2005).

\bibitem{Bernevig06}
B. A. Bernevig and S. C. Zhang, Phys. Rev. Lett. {\bf 96}, 106802 (2006).

\bibitem{Bernevig06_2}
B. A. Bernevig, T. L. Hughes, and S. C. Zhang, Science {\bf 314}, 1757 (2006).

\bibitem{Konig07}
M. K\"{o}nig et al., Science {\bf 318}, 766 (2007).

\bibitem{Schnyder08}
A. P. Schnyder, S. Ryu, A. Furusaki, and A. W. W. Ludwig, Phys. Rev. B {\bf 78}, 195125 (2008).

\bibitem{Kitaev09}
A. Kitaev, AIP Conf. Proc. {\bf 1134}, 22 (2009).

\bibitem{Qi09}
X.-L. Qi, Taylor L. Hughes, S. Raghu, and S.-C. Zhang, Phys. Rev. Lett. {\bf 102}, 187001 (2009). 

\bibitem{Hikihara08}
T. Hikihara, L. Kecke, T. Momoi, and A. Furusaki, Phys. Rev. B {\bf 78}, 144404 (2008).

\bibitem{Okunishi08}
K. Okunishi, J. Phys. Soc. Jpn. {\bf 77}, 114004 (2008).

\bibitem{Kolezhuk09}
A. K. Kolezhuk and I. P. McCulloch, Condens. Matter Phys. {\bf 12}, 429 (2009).

\bibitem{Chen14}
W. Chen and M. Sigrist, Phys. Rev. B {\bf 89}, 024511 (2014).

\bibitem{Katsura05}
H. Katsura, N. Nagaosa, and A. V. Balatsky, Phys. Rev. Lett. {\bf 95}, 057205 (2005).

\bibitem{Nogueira04}
F. S. Nogueira and K.-H. Bennemann, Europhys. Lett. {\bf 67}, 620 (2004).

\bibitem{Chen13}
W. Chen, P. Horsch, and D. Manske, Phys. Rev. B {\bf 87}, 214502 (2013).

\bibitem{Chen14_2}
W. Chen, P. Horsch, and D. Manske, Phys. Rev. B {\bf 89}, 064427 (2014).


\bibitem{Anandan89}
J. Anandan, Phys. Lett. A {\bf 138}, 347 (1989); {\bf 152}, 504
(1991).

\bibitem{Casella92}
R. C. Casella and S. A. Werner, Phys. Rev. Lett. {\bf 69}, 1625 (1992).

\bibitem{Lee94}
T. Y. Lee and C. M. Ryu, Phys. Lett. A {\bf 194}, 310 (1994).

\bibitem{Ryu96}
C. M. Ryu, Phys. Rev. Lett. {\bf 76}, 968 (1996).

\bibitem{Griffith89}
D. J. Griffith, "Introduction to electrodynamics", 2nd ed., Prentice-Hall (1989). 


\bibitem{Spielman00}
I. B. Spielman, J. P. Eisenstein, L. N. Pfeiffer, and K. W. West, Phys. Rev. Lett. {\bf 84}, 5808 (2000); Phys. Rev. Lett. {\bf 87}, 036803 (2001)

\bibitem{Kellogg04}
M. Kellogg, J. P. Eisenstein, L. N. Pfeiffer, and K. W. West, Phys. Rev. Lett. {\bf 93}, 036801 (2004). 

\bibitem{Tutuc04}
E. Tutuc, M. Shayegan, and D. A. Huse, Phys. Rev. Lett. 93, 036802 (2004). 

\bibitem{Eisenstein04}
J. P. Eisenstein and A. H. MacDonald, Nature {\bf 432}, 691 (2004).

\bibitem{Tiemann08}
L. Tiemann, J. G. S. Lok, W. Dietsche, K. von Klitzing, K. Muraki, D. Schuh, and W. Wegscheider, Phys. Rev. B {\bf 77}, 033306 (2008).

\bibitem{Huang12}
X. Huang, W. Dietsche, M. Hauser, and K. von Klitzing, Phys. Rev. Lett. {\bf 109}, 156802 (2012).


\bibitem{Sivan92}
U. Sivan, P. M. Solomon, and H. Shtrikman, Phys. Rev. Lett. 68, 1196 (1992).  

\bibitem{Seamons07}
J. A. Seamons, D. R. Tibbetts, J. L. Reno, and M. P. Lilly, Appl. Phys. Lett. {\bf 90}, 052103 (2007). 

\bibitem{DasGupta08}
K. Das Gupta, M. Thangaraj, A. F. Croxall, H. E. Beere, C. A. Nicoll, D. A. Ritchie, M. Pepper, Physica E {\bf 40}, 1693 (2008).


\bibitem{Kim11}
S. Kim, I. Jo, J. Nah, Z. Yao, S. K. Banerjee, and E. Tutuc, Phys. Rev. B {\bf 83}, 161401(R) (2011).

\bibitem{Kim12}
S. Kim and E. Tutuc, Solid State Commun. {\bf 15}, 1283 (2012). 

\bibitem{Gorbachev12}
R. V. Gorbachev, A. K. Geim, M. I. Katsnelson, K. S. Novoselov, T. Tudorovskiy, I. V. Grigorieva, A. H. MacDonald, S. V. Morozov, K. Watanabe, T. Taniguchi,
and L. A. Ponomarenko, Nature Phys. {\bf 8}, 896 (2012).

\bibitem{Berman12}
O. L. Berman, R. Ya. Kezerashvili, and G. V. Kolmakov, Phys. Lett. A {\bf 376}, 3664 (2012).

\bibitem{Aharonov84}
Y. Aharonov and A. Casher, Phys. Rev. Lett. {\bf 53}, 319 (1984).

\bibitem{Hirsch90}
J. E. Hirsch, Phys. Rev. B 42, 4774 (1990). 


\bibitem{Meier03}
F. Meier and D. Loss, Phys. Rev. Lett. {\bf 90}, 167204 (2003).

\bibitem{Schutz03}
F. Sch\"{u}tz, M. Kollar, and P. Kopietz, Phys. Rev. Lett. {\bf 91}, 017205 (2003).

\bibitem{Sun04}
Q.-f. Sun, H. Guo, and J. Wang, Phys. Rev. B 69, 054409 (2004).

\bibitem{Nagaosa08}
N. Nagaosa, J. Phys. Soc. Jpn. {\bf 77}, 031010 (2008).

\bibitem{Rashba82}
E. I. Rashba and M. D. Sturge, {\it Excitons}, North-Holland (1982).

\bibitem{Miller85}
D. A. B. Miller {\it et al.}, Phys. Rev. B {\bf 32}, 1043 (1985).

\bibitem{Kuga08}
S. Kuga, S. Murakami, and N. Nagaosa, Phys. Rev. B {\bf 78}, 205201 (2008).

\bibitem{Wilkens94}
M. Wilkens, Phys. Rev. Lett. {\bf 72}, 5 (1994).

\bibitem{Jackson99}
J. D. Jackson, "Classical electrodynamics", 3rd ed., John Wiley \& Sons (1999).

\bibitem{Castelnovo08}
C. Castelnovo, R. Moessner and S. L. Sondhi, Nature {\bf 451}, 42 (2008).

\bibitem{Morris09}
D. J. P. Morris {\it et al.}, Science {\bf 326}, 411 (2009).

\bibitem{Bramwell09}
S. Bramwell {\it et al.}, Nature {\bf 461}, 956 (2009).

\bibitem{Ryzhkin05}
I. A. Ryzhkin, JETP {\bf 101}, 481 (2005).







\end{thebibliography}
\end{document}